\documentclass[prl,twocolumn,showpacs]{revtex4}

\usepackage{bm}
\usepackage{graphicx}
\usepackage{amsbsy}
\usepackage{amsmath}
\usepackage{amsfonts}
\usepackage{amsthm}
\usepackage{color}
\usepackage{mathrsfs}
\usepackage{graphicx}% Include figure files
\usepackage{dcolumn}% Align table columns on decimal point
\usepackage{bm}% bold math

\def\ket#1{|\,#1\,\rangle}

\def\braket#1#2{\langle\, #1\,|\,#2\,\rangle}

\def\abs#1{|\,#1\,|}
\def\opone{\leavevmode\hbox{\small1\kern-3.8pt\normalsize1}}

\newcommand{\bwt}{\begin{widetext}}
\newcommand{\ewt}{\end{widetext}}
\newcommand{\bea}{\begin{eqnarray}}
\newcommand{\eea}{\end{eqnarray}}
\newcommand{\beq}{\begin{equation}}
\newcommand{\eeq}{\end{equation}}

\begin{document}

\title{State transformation in photon-echo quantum memory}

\author{A. Delfan Abazari$^1$, E. Saglamyurek$^1$, R. Ricken$^2$, W. Sohler$^2$, C. La Mela$^1$, and W. Tittel$^1$}
\affiliation{$^1$Institute for Quantum Information Science, University of Calgary, Calgary, Alberta, Canada T2N 1N4\\
$^2$Angewandte Physik, University of Paderborn, 33095 Paderborn, Germany}

%--------------------------------------------------------------------------------------------
% Abstract
%--------------------------------------------------------------------------------------------
\begin{abstract}
Quantum memory is a key element for quantum repeaters and linear optical quantum computers. In addition to memory, repeaters and computers also require manipulating quantum states by means of unitary transformations, which is generally accomplished using interferometric optical setups. We experimentally investigate photon-echo type atom-light interaction for the possibility to combine storage with controlled transformation of quantum states. As an example, we demonstrate unambiguous state discrimination of qubits and qutrits in an Ti:Er:LiNbO$_3$ waveguide cooled to 3K using states encoded into large ensembles of identically prepared photons in superposition of different temporal modes. The high robustness and flexibility of our approach makes it promising for quantum communication and computation as well as precision measurements.
\end{abstract}


%03.67.-a 	Quantum information
%32.80.Qk Coherent control of atomic interactions with photons
%42.50.Ex Optical implementations of quantum information processing and transfer in quantum optics
%78.47.jf Photon echoes
%03.67.Hk Quantum communication
%42.50.Ct 	Quantum description of interaction of light and matter; related experiments
%42.50.Gy 	Effects of atomic coherence on propagation, absorption, and amplification of light; electromagnetically induced transparency and absorption %42.50.Md 	Optical transient phenomena: quantum beats, photon echo, free-induction decay, dephasings and revivals, optical nutation, and self-induced transparency

\maketitle

%--------------------------------------------------------------------------------------------
% Introduction
%--------------------------------------------------------------------------------------------
Quantum memory \cite{Lvovsky2009} constitutes a key element for quantum repeaters \cite{Sangouard2009}, which promise to overcome the distance barrier of quantum communication such as in quantum cryptography \cite{Gisin2002}, as well as linear optical quantum computing (LOQC) \cite{Kok2007}. In addition to memory, repeaters and LOQC require unitary transformations, e.g. for single qubit rotations and controlled-NOT operations \cite{NielsenChuang}. To facilitate the identification of common performance criteria, e.g. thresholds for quantum error correction \cite{Shor1995}, quantum memory and state transformations can be described using the framework of \emph{gates}, with memory being the identity gate. However, their experimental realizations vastly differ. While quantum memory generally employs atom-light interaction \cite{Lvovsky2009}, all other gates required for quantum repeaters and LOQC currently rely on optical methods and possibly photo-detection \cite{Kok2007}.

In this letter, we demonstrate the possibility to combine storage and controlled transformation of light states by means of photon-echo based atom-light interaction. As an example, we demonstrate unambiguous state discrimination (USD) \cite{IvanovicDieksPeres} of qubit and qutrit states in an Ti:Er:LiNbO$_3$ waveguide cooled to 3K using states encoded into large ensembles of identically prepared photons in superposition of different temporal modes. This is an interesting task for many reasons: the impossibility to perfectly distinguish non-orthogonal states is at the heart of Quantum Information Processing \cite{NielsenChuang,Gisin2002,Scarani2004,Berlin2009}; USD requires state manipulation in Hilbert spaces of dimension exceeding two, which also enables to simplify quantum logic \cite{Lanyon2008} and to answer open questions related to entanglement and non-locality \cite{Gisin2008}; and the required multi-path interference underpins precision measurements.

\emph{Photon-echo quantum memory with multi-readout:}
Photon-echo quantum memory \cite{Tittel2009} relies on the interaction between light and an ensemble of atomic absorbers. The light to be stored is sent into the atomic medium with suitably prepared, inhomogeneously broadened absorption line. The preparation can be achieved by frequency-selectively transferring ions from the ground to some auxiliary state using optical pumping techniques \cite{pumping}, possibly followed by controlled broadening of the resulting absorption line through position dependent Stark or Zeeman shifts. After absorption, the excited atomic coherence rapidly decays as the atomic absorbers have different resonance frequencies. To recall the light, the initial coherence has to be re-established. This can be achieved by means of \emph{Controlled Reversible Inhomogeneous Broadening} (CRIB), which requires inverting the detuning of each atomic absorber with respect to the light carrier frequency \cite{Moiseev2001,Nilsson2005,Alexander2006,Kraus2006,Tittel2009}. Another possibility is to tailor the initial absorption line into an \emph{Atomic Frequency Comb} (AFC), resulting in re-emission of the absorbed light after a time that depends on the periodicity of the comb \cite{Hesselink1997,Afzelius2009,Riedmatten2008}.
In both approaches, the memory efficiency can theoretically reach 100\% \cite{Moiseev2001,Kraus2006,Hetet2008,Afzelius2009}.

The storage time in CRIB is limited by the homogeneous line-width of the optical transition or the width of the initially tailored absorption line \cite{Sangouard2007}. In AFC, it is predetermined by the spacing between the 'teeth' in the comb \cite{Afzelius2009}. For extended storage with on-demand recall, the excited atomic coherence can be temporarily transferred to coherence between two ground states, for which coherence times up to 30 seconds have been reported \cite{Fraval2005}. This transfer can be achieved by means of two $\pi$ pulses \cite{Afzelius2009b}, or via a direct Raman transfer \cite{Hetet2008b,Moiseev2008,Gouet2009,Hosseini2009}. 

Replacing the second $\pi$ pulse by several pulses of smaller area and variable phases, the input light is emitted in a coherent superposition of different temporal modes \cite{Moiseev2004}. This can also result in the interference of light from different input temporal modes in the same output mode. For two pulses, this atom-mediated coherent splitting/merging is equivalent to the combined action of an optical beam-splitter and a phase shifter, with temporal modes replacing spatial modes. As shown in \cite{Reck1994}, this enables implementing any discrete unitary transformation, and inspired our investigation.

CRIB was first demonstrated in 2006 with bright pulses of light \cite{Alexander2006}. Impressive progress followed for all photon-echo quantum memory approaches \cite{Hetet2008,Riedmatten2008,Hosseini2009,Afzelius2009b}. In particular, the beam-splitting operation has recently been implemented through tailoring of two atomic frequency combs instead of one \cite{Riedmatten2008}, and control fields of variable strength in a Raman-Echo Quantum Memory approach \cite{Hosseini2009}. Beam-splitting has also been reported via electromagnetically induced transparency \cite{Vewinger2007}.

Instead of CRIB or AFC, our investigation relies on the \emph{three-pulse photon echo} (3PPE) \cite{Mossberg1982}, which has been employed for storage and recall of sequences with up to 4000 bright pulses of light \cite{Lin1995}, and in information processing applications requiring phase coherent re-emission \cite{Arend1993,Kroll1993,Staudt2006}. The 3PPE derives from the two-pulse photon echo, in which a first, possibly modulated pulse of light (the \emph{data}), excites an inhomogeneously broadened atomic absorption line, and a second pulse reverses the subsequent dephasing of atomic dipoles \cite{AllenEberly}. In the 3PPE, this rephasing pulse is split into two pulses: the first (\emph{write}) pulse provides a time and phase reference for the data, and the second (\emph{read}) pulse provides the reference for the echo \cite{Mossberg1982}. This is similar to CRIB if the inversion of the atomic detunings is split into two parts: removal, followed by re-establishment with inverted sign.

If all pulse areas are sufficiently small, the amplitude and phase information encoded in the data is preserved in the 3PPE: $\varepsilon_{Echo}(t)\propto \int d\omega\varepsilon_1^*(\omega)\varepsilon_2(\omega)\varepsilon_3(\omega)\exp\{i\omega t\}$, where $\varepsilon_j(\omega)$ denote the Fourier transforms of the electric fields of the input and echo pulses, respectively \cite{Mitsunaga1992}. Assuming that the first pulse is the write pulse applied at t=0, the third is the read pulse applied at t=T, and that the power-spectra of both pulses are constant over the spectral width of the data, the electric field of the echo is given by $
\varepsilon_{Echo}(t)\propto\varepsilon_2(t-T)$. Hence, the signal replicates the data given by $\varepsilon_2(t)$ \cite{Mitsunaga1992}.

\begin{figure}
\begin{center}
\includegraphics[width=0.85\columnwidth]{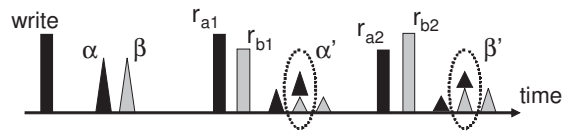}\\
  \caption{Read pulse sequence for the transformation in Eq. \ref{transformation}. The data consists of two pulses with electric field amplitudes $\alpha$ and $\beta$. The first read pulse, with amplitude $r_{a1}$ and phase $\phi_{a1}$, results in two echo pulses with amplitudes proportional to $e^{i\phi_{a1}}r_{a1}\alpha$ and $e^{i\phi_{a1}}r_{a1}\beta$. Similarly, the second read pulse creates two echo pulses with amplitudes $\propto e^{i\phi_{b1}}r_{b1}\alpha$ and $e^{i\phi_{b1}}r_{b1}\beta$. Provided the time difference between the two read pulses equals the difference between the data pulses, two echo pulses (one recalled by each read pulse) interfere, resulting in a field amplitude proportional to $e^{i\phi_{b1}}r_{b1}\alpha+e^{i\phi_{a1}}r_{a1}\beta$, i.e. the first component of the output state. The second component is created similarly using the second pair of read pulses. }
  \label{figure_readsequence}
  \end{center}
\end{figure}

A second echo, again resembling the data, can be triggered by a second read pulse. Hence, similar to using multiple control fields in photon-echo quantum memory protocols, the data is split coherently into two series of echoes separated in time, with splitting ratio and phase relation being determined by the read pulses. Moreover, 3PPE also allows the superposition of data \cite{Staudt2006} (see Fig. \ref{figure_readsequence}). Yet, despite the large similarity with genuine quantum memory protocols, the 3PPE approach is only suitable for storage of information encoded into large ensembles of identically prepared photons. This is due to an inherent amplification process when used to store few-photon pulses \cite{Ruggiero2009}. Nevertheless, 3PPE serves as an easily accessible test-bed to assess the validity of our joint approach to storage and transformation.

In 3PPE, each transformation is implemented using a particular sequence of read pulses. To find this sequence, we look at the transformation's matrix representation, associate amplitudes and phases of different pulses with its coefficients, and adjust the timing, i.e. the echo emissions, for the desired superposition. To illustrate this procedure, we consider the transformation

\begin{equation}
\hat{R} \left(
\begin{array}{c}
           \alpha \\
           \beta
         \end{array}\right)=\left(
  \begin{array}{c}
    e^{i\phi_{b1}}r_{b1}\alpha+e^{i\phi_{a1}}r_{a1}\beta \\
 e^{i\phi_{b2}}r_{b2}\alpha+e^{i\phi_{a2}}r_{a2}\beta \\
  \end{array}
\right)\equiv \left(
\begin{array}{c}
           \alpha ' \\
           \beta '
         \end{array}\right).
\label{transformation}
\end{equation}
\noindent
Two pairs of read pulses are required, each one generating an echo with amplitude proportional to one component of the output state (see Fig. \ref{figure_readsequence}). Adding read pulses, this multi-echo interference scheme is straightforwardly generalized to any transformation and to arbitrary numbers of input and output temporal modes, i.e. Hilbert spaces of arbitrary dimensions.

\emph{Unambiguous state discrimination:}
In classical physics, different states of signals are in principle perfectly distinguishable. In contrast, quantum physics allows the generation of non-orthogonal states that can only be distinguished in an imperfect way. For instance, using standard \emph{von Neumann} projection measurements, the attempt to identify the state is always conclusive, i.e. leads to a result, but the result is sometimes wrong. Assuming two pure input states $\ket{\varphi_{\pm}}$, the minimum error probability is given by the Helstrom (H) bound \cite{Helstrom1976}. For equal \emph{a priori} probabilities, it yields $P_e^{(H)}(opt)=\frac{1}{2}(1-\sqrt{1-\abs{\braket{\varphi_+}{\varphi_-}}^2})$.

Another approach to state discrimination is to drop the requirement of conclusiveness. This allows identifying non-orthogonal states with finite success probability unambiguously, i.e. without error. USD is implemented by embedding the quantum system under investigation into a  Hilbert space with additional degrees of freedom and unitarily evolving it in the higher dimensional space. This leads to an effective non-unitary transformation of the input state and can turn a set of non-orthogonal states into orthogonal ones, which can then be identified unambiguously. Assuming again two pure states with equal \textit{a priori} probabilities, the minimum probability for an inconclusive result is given by the Ivanovic-Dieks-Peres (IDP) bound \cite{IvanovicDieksPeres}: $P_?^{(IDP)}(opt)=\abs{\braket{\varphi_+}{\varphi_-}}$. It was later generalized to more than two states and dimensions, and arbitrary \textit{a priori} probabilities \cite{moreUSD}.

USD was first implemented by Huttner \textit{et al.} \cite{Huttner1996}, followed by Clarke \textit{et al.} \cite {Clarke2001}. Both experiments relied on polarization states of light, i.e. were limited to encoding non-orthogonal states in 2D (qubit) spaces. Mohseni \textit{et al.} extended these studies to one additional dimension \cite{Mohseni2004}. In the following, we demonstrate USD for linearly independent qubit and qutrit states encoded into large ensembles of equally prepared photons in superposition of different temporal modes. The demonstration employs 3PPE and is easily generalized to arbitrary dimension.

\emph{Experimental implementation:}
The experimental setup is similar to, and depicted in \cite{Staudt2006}. The cw 1532 nm line of an external cavity diode laser (Toptica) was amplitude and phase modulated using an acousto-optic modulator (AOM, Brimrose) to create write, data, and read pulses. The 15 ns long pulses were amplified by an erbium doped fiber amplifier (EDFA, Nuphoton Technologies), resulting in peak powers of $\sim$1 mW and 0.1 mW for the write pulses, and data and read pulses, respectively. The time difference between the write and the first data pulse was 300 ns, the data pulses were spaced by 100 ns, and the read pulse sequence was created $\sim$2 $\mu$s later. The light was sent into a 9 $\mu$m wide Titanium indiffused single mode waveguide, embedded in a 19 mm long, Z-cut, erbium diffusion doped LiNbO$_3$ crystal (surface concentration 3.6x10$^{19}$ cm$^{-3}$, optical depth at 1532 nm $\sim$three. For more information see \cite{Staudt2006,Baumann1997}). To inject and retrieve the light, we used standard telecommunication fibres that were butt-coupled against its front and end face using high precision translation stages (Attocube). The fibre-to-fibre coupling loss was $\sim$10dB. The Ti:Er:LiNbO$_3 $ waveguide was placed in a pulse tube cooler (Vericold), cooled to 3.1K, and exposed to a magnetic field of 0.4 T along its C$_3$-axis. To avoid heating of the sample and spectral hole burning due to amplified spontaneous emission from the EDFA, a second AOM, used as a shutter, was placed behind the EDFA. Transmitted pulses and echoes were detected at the output of the cryostat using a linear 1GHz-bandwidth photodetector (New Focus) and the electrical signal was stored on a 6GHz-bandwidth oscilloscope (LeCroy).
We used the transition between the lowest Stark levels of the Erbium $^{4}$I$_{15/2}$ and $^{4}$I$_{13/2}$ multiplets, which is characterized by an excited level lifetime of 2 ms and a coherence time under our experimental conditions of 18 $\mu$s. We note the importance of the magnetic field to reduce spectral diffusion \cite{Bottger2006}. The measurement sequence was clocked at 10Hz to avoid accumulation of information about the data in the excited state. The entire setup consisted of optical fibres and fibre pigtailed devices, which simplified alignment, resulted in high stability, and is promising for integration into future fibre-based quantum networks.

\begin{figure}[tt]
\begin{center}
\includegraphics[width=1\columnwidth]{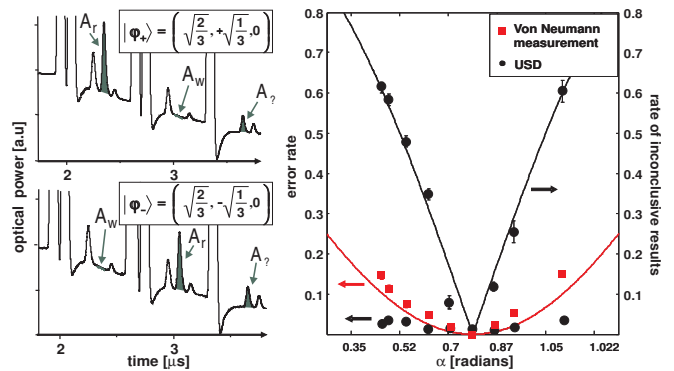}\\
  \caption{\emph{Left}: Averaged oscilloscope traces ($\sim$1000 measurements) showing read pulses (tops are cut) and echoes for input pulses with photons in states $\ket{\varphi_{\pm}}=(\sqrt{2/3},\pm 1/\sqrt{3},0)$. Each cluster of read pulses creates several echoes, similar to an optical interferometer. The echoes indicating the projections of interest are highlighted (other echoes indicate projections into auxiliary spaces). \emph{Right}: Rates for errors and inconclusive outcomes (points, squares) obtained from background-corrected echo areas, and H- and IDP-bounds for optimal implementation (solid lines). Experimentally obtained error rates slightly exceed the theoretical predictions, due to laser frequency fluctuations between data and read pulse creation, and small underestimation of the background.}
  \label{figure_results1}
  \end{center}
\end{figure}

We consider sets of two, generally non-orthogonal states $\ket{\varphi_\pm}=(\cos\alpha,\pm\sin\alpha,0)$,
$\alpha=[0,\pi /2]$. Following the procedure in \cite{Sun2001}, which minimizes the rate of inconclusive results, we first calculate the optimum output states $\ket{\varphi '_\pm}$ and find the transformation $\hat{U}$ that unitarily evolves $\ket{\varphi_\pm}$ into $\ket{\varphi '_\pm}=(\sqrt{2}\sin\alpha,0,\sqrt{\cos 2\alpha})$ or $(0,\sqrt{2}\sin\alpha,\sqrt{\cos 2\alpha})$, resp. (for $\alpha=[0,\pi /4])$. Knowing $\hat{U}$ then enables deriving the read-pulse sequence for the 3PPE.
The observation of an echo in a temporal mode encoding the first or second subspace unambiguously identifies the input state. Projections into the third subspace indicate an inconclusive result, i.e. the USD failed.

Fig. \ref{figure_results1} shows the experimental results. The almost perfect constructive or destructive interference for $\cos\alpha =\sqrt{2/3}$ (left hand picture) clearly demonstrates that USD is possible using a photon-echo approach. For a more quantitative assessment, we determined, for various angles $\alpha$, the background-corrected areas $A_i$ of the echoes in the three relevant temporal modes ($i\in \{w,r,?\}$ denotes \textit{wrong}, \textit{right}, and \textit{inconclusive}, resp.), and calculated the error rate $P_e=A_{w}/(A_{w}+A_{r})$ and the rate of inconclusive results $P_?=A_{?}/(A_{w}+A_{r}+A_?)$ for von Neumann and USD measurements \footnote{The background, which arose from detector response to the read pulses, was measured without data, i.e. echoes. We also normalized the echo areas using an experimentally determined scaling factor that reflects the effect of the read-pulses on the re-established atomic coherence.}. The error rate in USD remained roughly constant, below $\sim$4\%, while the error rate obtained with standard projection measurements increased up to $\sim$15\% with increasing overlap between $\ket{\varphi_\pm}$. Furthermore, the rate of inconclusive results rose for USD as $\alpha$ deviated from $\pi/4$. Our results agree well with the H- and IDP- bounds, reflecting the high stability of our implementation of interferometry.

\begin{figure}[tt]
\begin{center}
\includegraphics[width=1\columnwidth]{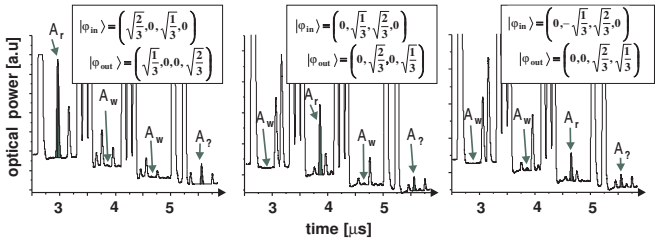}\\
  \caption{USD for three qutrit states. Each figure shows an oscilloscope trace averaged over $\sim$1000 measurements. The relevant output modes and echoes ($w,r,?$) are indicated.}
\label{figure_results2}
\end{center}
\end{figure}

To show the versatility of our approach, we extended the implementation of optimum USD to the case of three linearly independent qutrits (following again \cite{Sun2001}). The input and optimal output states, the read pulse sequence and the retrieved echoes are shown in Fig. \ref{figure_results2}. As before, the results agree well with theory. We emphasize the simplicity of this implementation, which only required adding three read pulses as compared to the previously discussed qubit case. This is a clear advantage to implementations based on standard optical interferometry \cite{Mohseni2004}, which quickly become impractical as the number of paths, i.e. the dimensionality of the Hilbert space required to process the data, increases.

\emph{Conclusion:}
The possibility to control the phase evolution of absorbers in an inhomogeneously broadened atomic ensemble after absorption of light allows not only storage of light states, but also inter-atomic state transformation. We have tested this idea with the example of transformations in up to 4D Hilbert spaces using a simple approach based on 3PPE and states of light encoded into large ensembles of identically prepared photons. The remarkable stability and versatility makes photon-echo based transformation an attractive alternative to standard optical interferometry, where the complexity of the setup rapidly escalates as the amount of interfering modes increases. While more investigations are required, we believe that our approach will find applications in quantum communication and LOQC as well as in high precision interferometry.
%%%%%%%%%%%%%%%%%%%%%%%%%%%%%%%%%%%%%%%%%%%%%%%%%%%%%%%%%%%%%
\acknowledgments     %>>>> equivalent to \section*{ACKNOWLEDGMENTS}
We thank Prof. Gisin for lending material, N. Sinclair and C. Simon for discussions, and V. Kiselyov for technical support. Financial support by NSERC, GDC, iCORE, QuantumWorks, CFI and AET is acknowledged.

%%%%%%%%%%%%%%%%%%%%%%%%%%%%%%%%%%%%%%%%%%%%%%%%%%%%%%%%%%%%%
%%%%% References %%%%%

\end{document}